\documentclass[12pt]{article}

\usepackage{amsmath} 
\usepackage{scicite}
\usepackage{makecell}
\usepackage{pgfplots}
\usepackage[]{geometry}
\usepackage{arydshln}
\usepackage{lineno}
\usepackage{hyperref}
\usepackage{times}

\newenvironment{sciabstract}{%
\begin{quote} \bf}
{\end{quote}}

\title{Revealing the unseen: Likely half of the Americans relied on others' experience when deciding on taking the COVID-19 vaccine}

\author{Azadeh Aghaeeyan$^{1\ast}$, Pouria Ramazi$^{1\ast}$, Mark A. Lewis$^{2}$\\
\\
\normalsize{$^{1}$Department of Mathematics and Statistics, Brock University, ON, Canada}\\
\normalsize{$^{2}$Department of Mathematics and Statistics and Department of Biology,} \\
\normalsize{University of Victoria, BC, Canada}\\
\\
\normalsize{$^\ast$To whom correspondence should be addressed; E-mail:  \{aaghaeeyan, pramazi\}@brocku.ca}
}

\date{}

\begin{document} 



\maketitle

\begin{sciabstract}
Efficient coverage for newly developed vaccines requires knowing which groups of individuals will accept the vaccine immediately and which will take longer to accept or never accept. 
Of those who may eventually accept the vaccine, there are two main types: success-based learners, basing their decisions on others' satisfaction, and myopic rationalists, attending to their own immediate perceived benefit. 
We used COVID-19 vaccination data to fit a mechanistic model capturing the distinct effects of the two types on the vaccination progress.
We estimated that $47\%$ of Americans behaved as myopic rationalists with a high variation across the jurisdictions, from $31\%$ in Mississippi to $76\%$ in Vermont. 
The proportion was correlated with the vaccination coverage, proportion of votes in favor of Democrats in 2020 presidential election, and education score.

\end{sciabstract}

Vaccination is a primary measure to reduce morbidity and mortality of new infectious diseases \cite{lin2022effectiveness}, especially in areas that admit high coverage \cite{whoTarget}. 
High vaccination coverage requires high vaccine acceptance--a collective outcome of individual-level decision-making processes \cite{WHOhesitancyFactors}. 
Particularly, when facing a new vaccine, people need to decide between the promised immunity followed by potential side-effects associated with accepting the vaccine versus the mortality risk and governmental restrictions of refusing the vaccine. 
Individuals, however, vary in their decision-making strategies \cite{molleman2014consistent,toelch2014individual,mesoudi2011experimental}, and different strategy compositions can result in different collective outcomes \cite{van2007new}, including different vaccination rates and coverage \cite{ndeffo2012impact,CDCtracker}.
Effective promotion of vaccination benefits from
identifying the different types of decision-makers and deploying tailored communication methods.

In various contexts, people are known to be mainly one of the following two decision-making types \cite{bass1969new,tanny1988innovators,van2007new,mesoudi2011experimental}: 
\emph{(i)} those who learn from others' success, particularly, those who take the action of others with a higher payoff, known as \emph{success-based learners} \cite{mcelreath2008beyond,van2015focus}, or \emph{imitators} \cite{PINGLE1995281}, and 
\emph{(ii)} those who base their decisions on their own perceptions of the environment, often aiming to maximize their instant perceived payoff \cite{toelch2014individual, mesoudi2011experimental}, named \emph{myopic rationalists}, or \emph{best-responders} \cite{KANIOVSKI200050}.
Exclusive populations of myopic rationalists who perceive the social context similarly are likely to reach satisfactory decisions \cite{ramazi2016networks} whereas exclusive populations of success-based learners may undergo perpetual changes of decisions \cite{ramazi2022lower,van2015focus}.
Mixed populations of the two may exhibit a wide range of behaviors depending on the proportions \cite{ndeffo2012impact,le2020heterogeneous}.
Understanding the collective outcome of mixed populations, hence, requires knowing the proportions of success-based learners and myopic rationalists. 

The two types of decision makers also differ in the type of information they attend to \cite{van2015focus}. 
Myopic rationalists seek information that shapes their payoffs, whereas success-based learners focus on the satisfaction achieved by others.
Hence, in addition to providing insights to the vaccination dynamics, knowing 
the proportions of the two types  
inform health management and media. 

The proportion, however, is unknown and needs to be estimated from data -- a yet unaccomplished challenging task.
The challenge is mainly because the proportions of decision-makers are unobserved. A natural approach to tackle this problem is to build a mechanistic model with the proportion of decision-makers as a parameter and estimate it by fitting the equations to existing data on measured quantities such as vaccine uptake. 
However, for this approach to be successful, it must be first shown that such a parameter is identifiable.
 Several other challenges exist. There may be insufficient detail in data from previous seasonal or relatively old diseases, where decisions were made typically once a year  \cite{moore2021vaccination,FDA} or lifetime \cite{CohenWaning}. Baseline trust regarding a vaccines’ effectiveness and safety may change as an initially distrustful public gradually becomes accustomed to it \cite{beleche2021covid}. The resulting hesitancy followed by the desire for immunity results in frequent decision revisions, e.g., on a bi-weekly basis \cite{beleche2021covid}.
 Finally, not all the population may be concerned about the vaccine as can be seen in some previously recorded vaccination programs \cite{fluVacDash}. 


The recorded data on COVID-19 in the US provides a unique opportunity to tackle each of these challenges. 
Although the proportion is unobservable, the two decision-making groups have different vaccination paces and differently shape the vaccination progress curve, which is observable. 
Indeed, we prove that the proportion of myopic rationalists is an identifiable parameter.
Hence, by fitting the collective decision-making dynamics of the two types to vaccination data, we can estimate the proportion and use bootstrapping methods to obtain confidence intervals.
Owing to the importance of timely vaccination and thanks to the advancement of technology, the vaccination data are available on a daily basis \cite{hale2021global}. 
The changes in baseline attitude toward the vaccine's side effects were captured through longitudinal surveys \cite{survey}.
Moreover, almost every resident had to decide on vaccination \cite{fda-6months}.

Our objective was to estimate the proportions of myopic rationalists and success-based learners in each of the 50 states of the US and the District of Columbia (D.C.) separately, in deciding whether to take first dose of COVID-19 vaccination.
We excluded later doses as they are influenced by the experience in the first dose \cite{karaivanov2022covid}.
We developed a  mechanistic model to capture the behavior of the two types of interacting individuals augmented by a third group \emph{vaccine refusers}, those \textit{a priori} known to refuse the vaccine based on surveys \cite{hesitancy}.
A fixed parameter ($\alpha$) was used as the proportion of myopic rationalists in the population in each jurisdiction.
The perceived excess payoff of vaccination is shaped by the epidemiological conditions represented by weekly cases and deaths, the risk of vaccine side effects, and the social and governmental pressure on unvaccinated individuals.

\section*{Results}
Our fitting results suggested that $47\%$ ($\alpha = 0.47$) of Americans aged $12$ years and above behaved as myopic rationalists in receiving the first dose of COVID-19 vaccine, i.e., about $131$ millions out of $279$ million Americans, an equal proportion acted as success-based learners, and the remaining $6\%$ were COVID-19 vaccine refusers.
The estimated percentage of rationalists across the US varied from  $31\%$ in Mississippi to $76\%$ in Vermont (Fig. 1). 
In fifteen states, myopic rationalists composed more than $50\%$ of residents.

To ensure that the estimations were as robust as possible, we reported the confidence intervals (CIs) obtained from non-parametric bootstrapping as they resulted in larger CIs than parametric bootstrapping (Table S1).
The length of $95\%$ CI of the proportion of myopic rationalists  was  $0.13$ or less for all jurisdictions (Table S1). 
For 20 states, the upper limit of the CI was lower than the estimated proportion across the US, i.e., $0.47$, as well as the estimated proportions of 25 other states, implying a significant difference in the proportions. 
Particularly, the proportions of 46 states including California and Florida fell between the upper confidence limit of Mississippi, the state with the lowest proportion  and the lower confidence limit of Vermont, the state with the greatest proportion
(Fig. 2).

As of November 2021, all eligible American myopic rationalists were vaccinated.
Success-based learners had  lower vaccine coverage.
While in all jurisdictions more than half of the imitators received at least one dose of a COVID-19 vaccine, in no jurisdiction were they all vaccinated.

The weekly number of vaccinated individuals, particularly rationalists, was greatly influenced by the vaccine doses available during the first months of vaccine rollout which was due to vaccine supply limitations (Fig. 3A, B, D, E).
The changes in the perceived benefit of vaccination in some states were negligible compared to other states (Fig. 3C, F, Figs. S2-S26).  

The proportion of myopic rationalists was correlated with the vaccination coverage, proportion of votes in favor of Democrats in the 2020 presidential election, and education score (Table 1--see Table S2 for the correlation with the other  explanatory variables). The proportion of myopic rationalists was not highly correlated with the accessibility to the vaccination facilities (Table 1--last two rows).
We evaluated the impact of each model component on the fitting results.
The averaged residual sum of the squares (RSS) over all jurisdictions increased if any of the components were altered. 
More specifically, by assuming a fully myopically rational population ($\alpha = 1$), the RSS increased by $616\%$, whereas a population of success-based learners ($\alpha = 0$), resulted in an increase of $626\%$.
To investigate the effect of the possible delay in delivering the vaccine doses compared to what was reported in the data, we introduced a 3-day delay based on California vaccine distribution plan \cite{Cadence}.
The estimated composition of myopic rationalists and success-based learners at the state level was fairly insensitive to the introduced delay. More specifically, the variations in the estimated proportion of myopic rationalists were less than $10\%$(see Table S3). 
\begin{table}
\centering
\caption{Linear correlation between explanatory variables and the estimated proportion of myopic rationalists. }
\begin{tabular}{lrrr}
\textbf{Predictor variable}  & \textbf{Pearson-r} & \textbf{R-squared} \\
\hline
\makecell[l]{Vaccination coverage}  & $0.87$ & $0.76$\\ \hline
 \makecell[l]{Proportion of votes in favor of Democrats} & $0.82$ & $0.68$\\ \hline
\makecell[l]{Education score} & $0.74$& $0.54$ \\   
\hline \hline
    \makecell[l]{Proportion of people living further than \\10 miles from vaccination facilities} & $-0.37$ & $0.14$\\ 
    \hdashline[0.5pt/5pt] 
    \makecell[l]{Density of vaccination facilities} & $-0.24$ & $0.06$\\
\end{tabular}
 \label{tab:correlation}
\end{table}

\subsection*{Discussion}
We considered vaccination coverage as a collective outcome resulting from decisions of individuals that are known to be mainly either myopic rationalists or success-based learners. 
Although crucial to the vaccination dynamics, the proportion of the two types of decision makers has to date been unknown and not measured directly. 
We tackled this problem by developing
a mechanistic model capturing the evolution of vaccinated population with fixed ratios of the two types of decision-makers in the context of vaccine uptake.
We additionally considered a third type of vaccine refusers who never intended to be vaccinated. 
By fitting the model to data on the number of first-dose COVID-19 vaccinated individuals in the US, we found that, excluding the $6\%$ vaccine refusers, about half of the Americans behaved as myopic rationalists, and half as success-based learners.  
The results may inform health management and guide tailored communication towards promoting vaccination uptake.

The main challenge in this work was how to estimate this unobserved proportion of myopic rationalists.  
We took the natural approach of building a mechanistic model with the proportion of decision-makers as a hidden parameter and then estimated it by fitting the equations to existing data on observed quantities such as vaccine uptake. 
However, there are two potential drawbacks with this approach. 
First, there could be parameter identifiability issues, resulting in a low estimation confidence.
We tackled this issue by proving that the proportion of myopic rationalists is an identifiable parameter that can be uniquely estimated. 

The fact that the proportion is identifiable implies that it can be estimated using a sufficiently rich dataset, which appeared to be the case for our study due to the sufficiently narrow confidence intervals. 
As the fitting errors did not follow a normal distribution and were temporally correlated, the confidence intervals could not be obtained based on the common assumption of independence and normally distribution. 
We addressed this issue by using non-parametric bootstrapping methods and applying an auto-regression to reduce the temporal correlation.

The second drawback is that even with a unique solution, there may be other possible models for the observed human behaviour. 
Alternative explanations of the vaccine uptake trends could be a population consisting of \emph{(i)} multiple groups of success-based learners with different learning rates, \emph{(ii)} multiple groups of myopic rationalists with different uptake rate, and \emph{(iii)} multiple groups of both success-based learners learning at different rates and myopic rationalists with different uptake rates who are interacting. 
The first two hypotheses, however, ignore the evidence from other studies of the coexistence of success-based learners and myopic rationalists.
Indeed,  the dichotomy of decision-making populations has been acknowledged in previous vaccination studies \cite{ndeffo2012impact,dinterplay} as well as other contexts, such as marketing, psychology, and cultural evolution \cite{bass1969new,mahajan1990new,tanny1988innovators,van2007new,mesoudi2015higher, metz2010spatial}.
A population with multiple groups of both decision-makers is likely to be more realistic.
Despite this, we have considered the simplest model that includes both myopic rationalists and success-based learners and the more complex models are left for future work.

The proportion of the two types of decision-makers has not been estimated in previous vaccination studies; 
only homogeneous models have been fitted to data \cite{bauch2012evolutionary,mills2022effect}.
Some marketing studies estimated the relative proportion of the two types in adopting several products using the Bass model \cite{van2007new,cavusoglu2010information, nia2016social}. 
Our work complements these studies in that \emph{(i)} rather than estimating the proportion among the final adopters, we estimated the proportion in the whole population, including unvaccinated individuals, \emph{(ii)} we considered the influence of time-varying variables including epidemiological indices, and the perceived risk associated with vaccine side effects, \emph{(iii)} we incorporated supply limitation in the model, and \emph{(iv)} we showed that the proportions of these decision-makers are identifiable.




We found that myopic rationalists greatly determined the evolution of the number of weekly vaccinated individuals (vaccination speed) during the first months of the vaccine roll-out -- a crucial factor in the success of vaccination programs \cite{kim2022balancing,smith2021constitutes,reddy2021clinical}. 
In later months, the vaccination speed was mainly determined by success-based learners. 
Considering the excess payoff formulation, the only factor capable of inducing a negative excess payoff is the perceived risk associated with side effects. This factor, however, according to a longitudinal survey,  declines over time \cite{beleche2021covid}. 
This suggests that once vaccination has a higher payoff than remaining unvaccinated, which was the case during the first months, all myopic rationalists tended to take the vaccine--the only limiting factor being vaccine availability.
Although success-based learners also tended to take the vaccine once the excess payoff is positive, they based their decisions on others' success, determined by the magnitude of the excess payoff and the number of vaccinated individuals.
Hence, they tend to delay vaccination during the first months of the vaccine roll-out but exhibit a higher vaccination speed when many are vaccinated and available to imitate.

Myopic rationalists also greatly determined the vaccination coverage -- another key factor in the success of vaccination programs \cite{christie2021guidance}.
Our findings suggest that as of November 2021, all American rationalists have received their first dose of vaccine and consequently, excluding the vaccine refusers, those who remained unvaccinated were success-based learners.

The results suggest the proportion of success-based learners as a factor behind the collective vaccination behavior and contributing to the success of mass vaccination programs. 
The beginning of vaccination programs may mainly address access issues to expedite vaccination among myopic rationalists. 
Later on, vaccine-promoting interventions should be tailored to success-based learners to increase their perceived benefit of vaccination. 
The initiation of this second phase depends on the composition of the decision-makers, i.e., the more success-based learners in the population, the sooner the second phase should be initiated.
 


We observed high variations in the estimated proportion of myopic rationalists across the US jurisdiction, ranging from $0.31$ to $0.76$. 
The result is in line with studies highlighting the variation of different indices throughout the states \cite{innovationStates,falcettoni2020comparison}.
The delivered doses in the first months shaped the number of vaccinated rationalists.
However, the variation may not be attributed to vaccine supply disparity, because the doses were distributed proportionally to the state populations \cite{distribution}.


There are candidate factors to explain the inter-state differences in the ratio of myopic rationalists and success-based learners.
The proportion of myopic rationalists was  positively correlated with the vaccination coverage (Pearson-r = $0.87$).
The proportion of myopic rationalists was also positively correlated with the proportion of votes in favor of Democrats in 2020 presidential election (Pearson-r = $0.82$).
These findings are then consistent with the known relation between vaccination coverage and 2020 presidential election outcome \cite{kff1}.
The ratio of myopic rationalists was also positively associated with education score.
In view of the correlation between the ratio of myopic rationalists and vaccination coverage, this is indirectly corroborated by studies suggesting American adults with less education being less likely to receive the vaccine \cite{nguyen2021covid,beleche2021covid}. 

  

Our model shows how interactions between myopic rationalist and success-based learner groups have determined the course of the vaccination for COVID-19 in the United States. Admittedly, true population could have additional heterogeneities, where different age, gender, or socio-economic status groups have different perception of the payoff \cite{ndugga2021latest,saelee2022disparities,WHO}. 
The extension of the model to capture this heterogeneity is subject to future work.  
Another limitation of this work comes from assuming a well-mixed population rather than the more realistic structured population as supported by Twitter echo-chambers \cite{cossard2020falling}. 
This assumption may not greatly affect myopic rationalists as important information about the excess payoff is often obtained from and shared by all reputable and publicly accessible news agencies. In the same way, online influencers -- usually perceived as the most successful -- make the interactions of success-based learners more well-mixed than structured \cite{smallerWorld}. 
 Overall, while acknowledging the aforementioned limitations, the results of this study provide strong evidence that the COVID-19 vaccine uptake dynamics were determined by interactions between myopic rationalists and success-based learners and at least half of the Americans relied on the success of others in taking COVID-19 vaccine.
\begin{figure}
\centering 
\includegraphics[trim=2.4cm 1.4cm 0.8cm 3cm, clip=true,width=11.4cm]{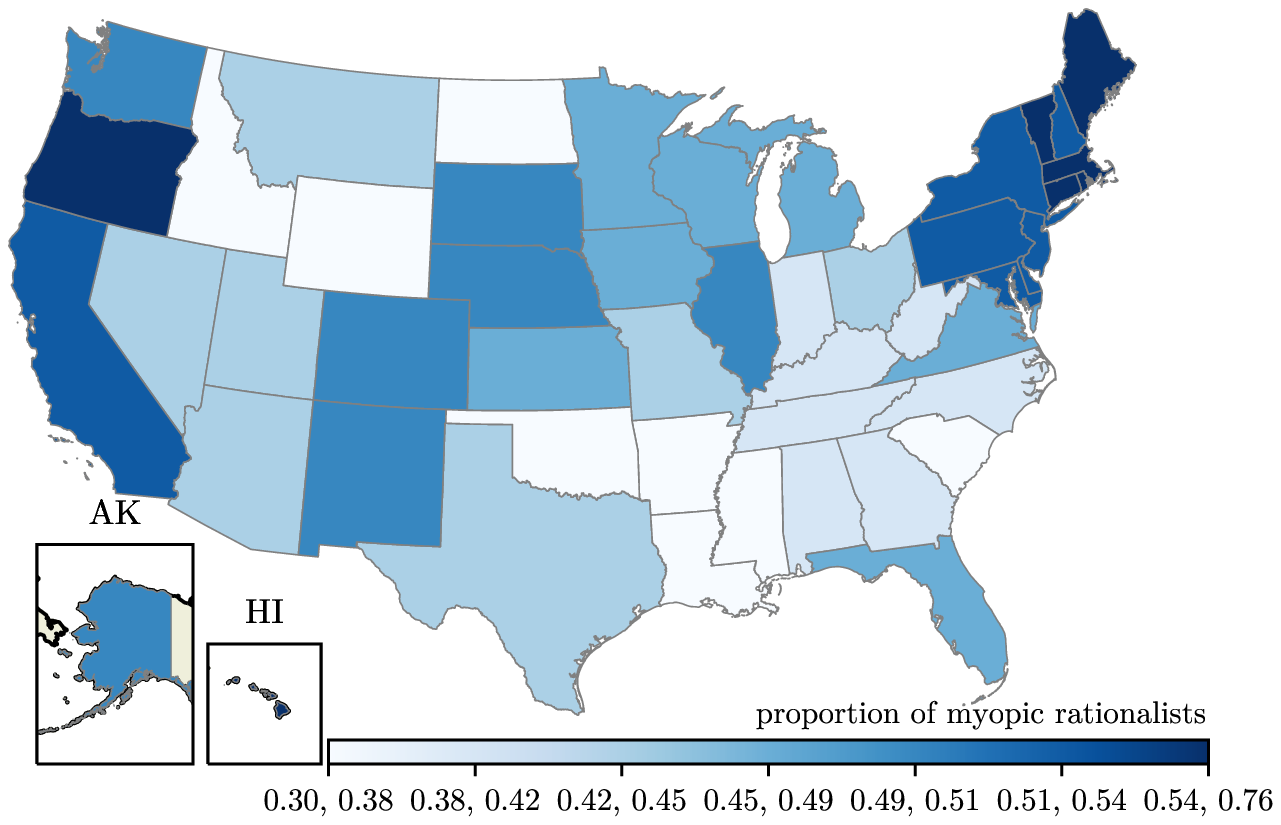}
\caption{\textbf{Estimated proportion of myopic rationalists in taking the first dose of COVID-19 vaccination across the 50 states and the D.C in the US}. 
Darker colors show a higher proportion of myopic rationalists.
For each jurisdiction, the proportion of myopic rationalists aged  12 years and above was estimated by fitting the developed mechanistic model to the data on weekly newly vaccinated individuals starting from Dec 2020 to Nov 2021. 
The vaccine supply limitations during the first months of vaccine roll-out were captured by using the data on delivered doses to each jurisdiction over time.
The national wide estimated proportion of myopic rationalists was  $0.47$. 
There was a high degree of variation across the 51 jurisdictions, i.e., $0.31$ for Mississippi to $0.76$ for Vermont.
About $52\%$ of the residents of the states located in the Northeast region behaved as myopic rationalists in taking COVID-19 vaccination, while in southern states, this proportion was about $43\%$. The estimated proportion of myopic rationalists in West and Midwest regions was respectively $0.49$ and $0.46$.
}\label{fig:map}
\end{figure}

\begin{figure}
\centering
\includegraphics[width=.8\linewidth]{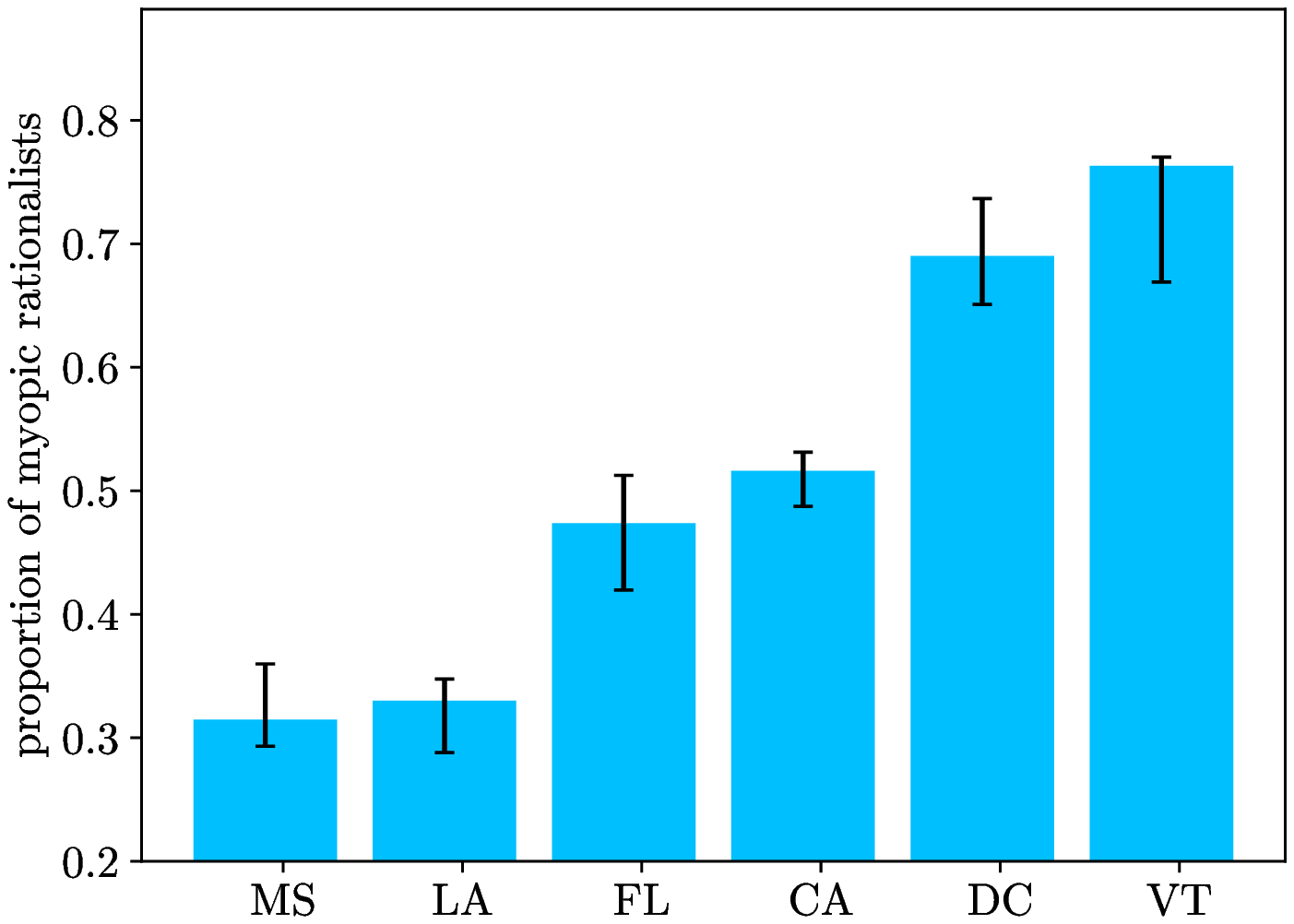}
\caption{ \textbf{Estimated $\%95$ confidence interval of the proportion of myopic rationalists for five states of the US and the D.C.} 
The US of Vermont and has the highest proportion of myopic rationalists, while Mississippi has the lowest.
Error bars indicate the $\%95$ confidence interval obtained using non-parametric bootstrapping. 
For all states and the D.C., the lengths of the confidence intervals were $0.13$ or shorter.
}
\label{fig: alphaBar}
\end{figure}

\begin{figure*}
\centering
\includegraphics[trim=1.5cm 0.7cm 1.3cm 2.5cm, clip=true, width = 17.8cm]{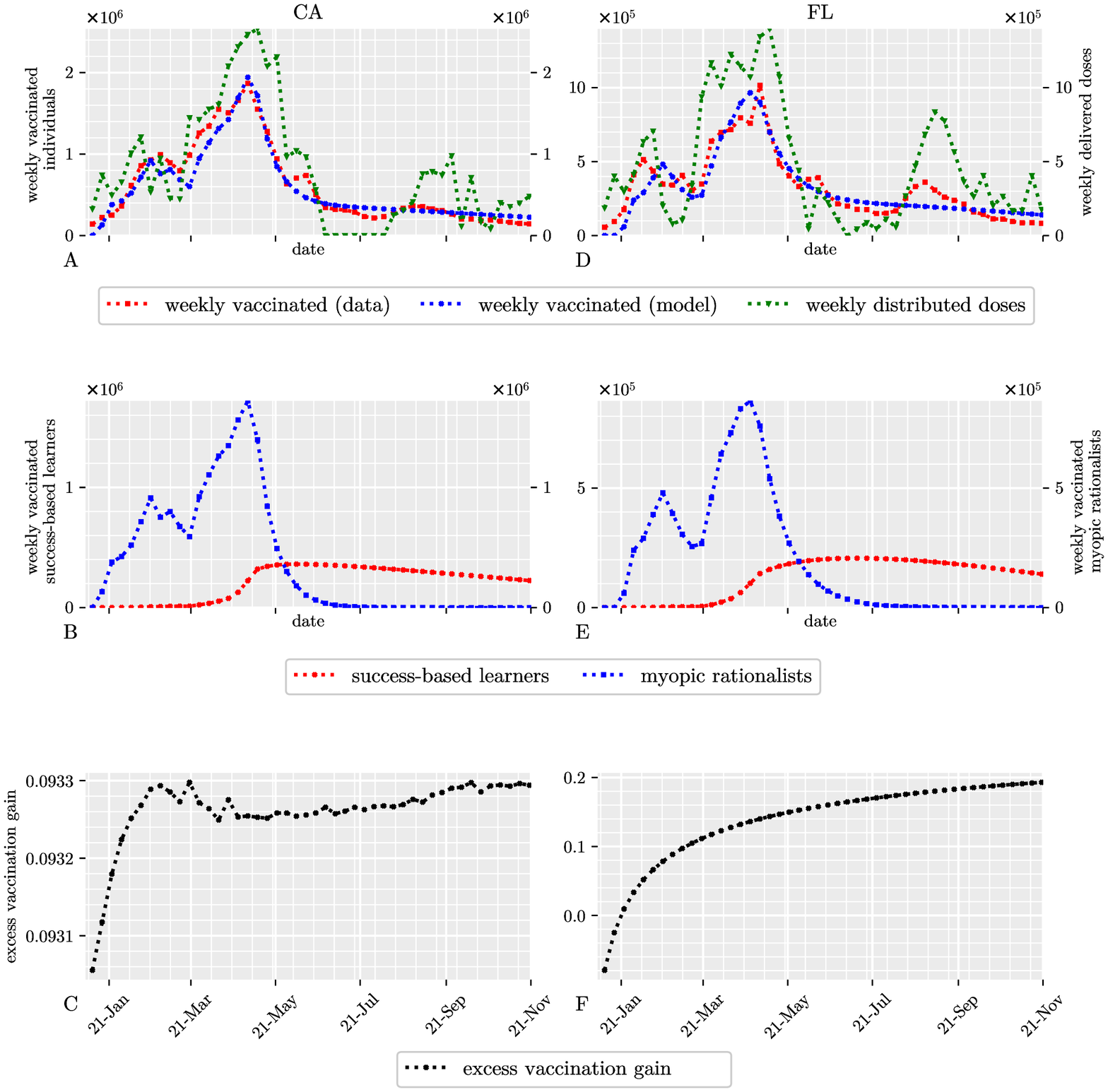}
\label{fig:temporalGraph}
\end{figure*}
\begin{figure} [t!]
 \caption{\textbf{Number of weekly vaccinated individuals, myopic rationalists, success-based learners, and the excess payoff over time for states of California and Florida, Dec 2020-Nov 2021}).
Each panel includes 46 data-points.
Left panels depict the results and data for California and the right ones are for Florida. 
Panels A and D depict the weekly number of vaccinated residents in red, its estimation in blue, and weekly number of delivered doses in green.
Panels B and E depict the estimated weekly number of vaccinated myopic rationalists and success-based learners in blue and red, respectively.
The panels in the last row depict the evolution of the benefit of vaccination (excess payoff).
The number of weekly vaccinated individuals is shaped by the vaccine doses available particularly during the first months of vaccine roll-out (Panels A and D).
Almost all myopic rationalists receive the vaccine by July, 2021 (the blue curves in Panels B and E).
As vaccination progresses,  success-based learners are persuaded to receive a vaccine  (Panels B and E).
Over the studied time interval, the perceived benefit of vaccination among Californians changes slightly compared to that of Floridians (Panels C and F).}
\end{figure}
{
\section*{Materials and Methods}
\subsection*{Data}
 To calculate the cumulative number of available COVID-19 vaccine doses for administering the first shot,  we used the temporal data on the number of delivered doses to each jurisdiction's provider locations. The data were available on the website of Centers for Disease Control and Prevention  (CDC) \cite{CDC3} and captured the last part of the vaccine distribution process \cite{distributionDef}. 
  To avoid double-counting, the numbers of individuals completing the primary series of a two-dose vaccine, receiving the first and the second booster doses were subtracted from the data. 
    In all $51$ jurisdictions across the US, the decision whether to vaccinate children under the age of 12 was left up to their guardians. Hence, we narrowed the population to those of 12 years and older. As the emergency authorization of vaccinating children of age 5 through 11 was issued on late October, 2021,
   we set October 30, 2021 as the last day of the dataset \cite{fda-5years}.
  
  December 14, 2020 was selected as the starting date.
  Daily data on the total number of individuals who received their first dose, the number of individuals who completed the primary course of vaccination, and the number of individuals who received their first and second booster doses were collected from \cite{CDC1}. 
  State level data on daily new cases and deaths were obtained from 
  \cite{CDC2}.
The negative values were replaced with zero and then  we replaced zero new cases with their immediate non-zero values in the next day. 
  We converted the temporal scale from daily to weekly. 
  The state-wide estimated proportion of COVID-19 vaccine refusers was obtained from 
  \cite{hesitancy} (see SI).
  
  To estimate the trend of reduction in perceived risk of vaccine side-effects, we used the data collected by the US Census Bureau \cite{survey}. 
  To determine the start date at which containment, closure or health policies differentiated based individuals' vaccination status, we used the US sub-national data available in
  \cite{hale2021global}. 
  The data on state level population by age was obtained from \cite{subpopulation}. 

  We additionally explored a number of features that could possibly relate to the proportions of success-based learners and rationalists. 
  We collected data on votes in favor of each political party from \cite{presidential}, income per capita and proportion of people below the poverty line from \cite{SVI}, state-level education score from \cite{educationalstore}, proportion of population with driving distance greater than 10 miles to the closest vaccination facility and density of potential vaccination facilities from \cite{berenbrok2021access,westhealth}, age, sex distribution, education level distribution, number of intensive care unit beds, ventilator capacity and percent insured residents, meat plants, religious congregation ratio and immigrant student ratio from \cite{haratian2021dataset}.
\subsection*{Model Formulation}
For each jurisdiction, we considered a large fixed population of $N$ interacting individuals of which an unknown fixed portion $\alpha$ are myopic rationalists, a known fixed portion (refer to SI, Section Data) are COVID-19 vaccine refusers and the remainder are success-based learners.
The individuals decide based on the conceived excess payoff of receiving a vaccine defined by

\begin{align} \label{eq:excesspayoff}
 \Delta\pi (t) = C_{s}(t) + {C}_d D(t)/N + {C}_i I(t)/N,
 \end{align}
where $I(t)$ and $D(t)$ are the weekly number of confirmed cases and deaths from COVID-19 as a function of time. 
The conceived excess payoff of vaccination comprises of three terms: 
First, the quantity $C_s(t)$ is the perceived benefit of vaccination in the absence of confirmed cases or deaths and is equal to the perceived socio-economic benefit of vaccination, $c_{\bar{v}}$  minus the perceived risk of vaccine associated side-effects, $C_v(t)$. 
Second, the perceived vaccine-induced risk reduction of dying from COVID-19, $C_d$, that is the perceived cost of dying, times the perceived effectiveness of the first dose, multiplied by $D(t)/N$, which is the perceived chance of death from COVID-19. 
Third, the perceived vaccine-induced risk reduction of COVID-19 infection is defined similarly as $C_i I(t)/N$.
According to Household Pulse survey, a longitudinal survey conducted by Census Bureau, the perceived risk of vaccine associated side-effects declines over time \cite{beleche2021covid}. Our fitting result indicates that this decreasing trend can be best described by a power law function of the form $(t-t_0)^{\lambda}$ (see SI). Hence, $C_v(t)$ is replaced with $c_{v_{0}}(t-t_0)^{\lambda}$ where $c_{v_{0}}$ is a free parameter (see SI).
The perceived socio-economic benefit of vaccination is modeled by a free parameter, $c_{\bar{v}}$. 

The myopic rationalists follow the \emph{best-response dynamics} \cite{sandholm}, that is, they compare the payoff to vaccination with that of remaining unvaccinated and choose to get vaccinated if the excess payoff is positive, i.e.,  $\Delta \pi(t)>0$. 
The success-based learners, however, base their decisions on their interactions with others in the population: Upon an interaction, they compare their own payoff with those of others.
If the others have a higher payoff, the success-based learner imitates their decisions with a probability proportional to the excess payoff. 
The individuals may update their decisions over vaccination independently at an exponential rate, or over discrete periods of time. In either case, the collective decision making process can be approximated by the mean dynamics \cite{sandholm}, which we do in \eqref{eq:lvdot} and \eqref{eq:mvdot}.
Focusing on the first dose, the vaccinated people never change their strategies. 

Let $M_v(t)$ (resp. $L_v(t)$) denote the number of vaccinated myopic rationalists (resp. success-based learners). 
Similarly, define $M_{\bar{v}}(t)$ (resp. $L_{\bar{v}}(t)$) as the number of unvaccinated myopic rationalists (resp. success-based learners). 
Let $M_s(t)$ (resp. $L_s(t)$) denote the number of vaccine-seeker myopic rationalists (resp. success-based learners) that is those who are unvaccinated and want to receive a dose of vaccine. The  number of vaccine-seeker myopic rationalists (resp. success-based learners)  will be
$M_s(t)=  M_{\bar{v}}(t) \times {\rho^M(t)}$ (resp. ${{L}_{s}(t)}=  {L}_{\bar{v}}(t) \times {\rho^L} (t)$),
where $\rho^M(t)$ (resp. $\rho^L(t)$) denotes the proportion of unvaccinated myopic rationalists (resp. success-based learners) who are vaccine-seekers, which is influenced by the number and payoffs of vaccinated and unvaccinated individuals. 
Inspired by \cite{sandholm}, the proportion of success-based learners who are vaccine seeker is 
    $\rho^{L}(t)=\eta_{v}(t){N_{v}(t)}/{N}$, where
$N_v(t)$ denotes the cumulative number of vaccinated individuals up to time $t$, which is equal to $L_v(t) + M_v(t)$. 
Therefore, $N_v(t)/N$ denotes the probability that an unvaccinated success-based learner meets a vaccinated individual. 
The term $\eta_{v}(t)$ is proportional to the probability of being a vaccine-seeker after interacting with a vaccinated individual. 
A pairwise proportional comparison form is considered for $\eta_{v}$, i.e., 
$\eta_{v}(t)= \sigma[\Delta \pi]_+$, where $[x]_+$ equals $x$ if $x>0$ and equals zero otherwise, and $\sigma$ is the constant of proportionality \cite{WANG20161}. 
In other words, if the vaccinated individuals' payoff is higher, i.e., $\Delta \pi(t) > 0$, the success-based learners will be vaccine-seeker with a probability proportional to $\Delta \pi(t)$. 
 The proportion of vaccine-seeker rationalists can be written as
$\rho^M= H(\Delta \pi)$,
where $H(\cdot)$ is the Heaviside function, which equals one for a positive argument and zero otherwise.
 The number of non vaccine-refusers is denoted by $N_n$, which equals to the total non-hesitant population aged 12 years and above.
 The number of success-based learners then equals $(1-\alpha_1)N_n$, where $\alpha_1 = \alpha N/N_n$. 
 By subtracting the number of vaccinated success-based learners $L_v$, we obtain the number of unvaccinated success-based learners, $L_{\bar{v}} $. Similarly, the number of unvaccinated myopic rationalists, $M_{\bar{v}}$, equals $\alpha_1 N_n$ from which the number of vaccinated rationalists $M_v$ is subtracted.
The number of vaccine-seeker success-based learners at time $t$ is then

 \begin{align}  \overbrace{L_s(t)}^{\substack{\text{number of }\\ \text{vaccine-seeker} \\ \text{success-based}\\\text{learners}}} = \overbrace{((1-\alpha_1)N_n-L_v(t))}^{\substack{\text{number of}\\\text{ unvaccinated }\\ \text{success-based learners}}}\overbrace{\frac{L_v(t) + M_v(t)}{N}}^{\substack{\text{proportion of}\\ \text{vaccinated} \\ \text{individuals}}}\overbrace{\sigma [\Delta \pi(t)]_+}^{\substack{\text{dimensionless}\\
   \text{excess}\\\text{payoff}}}.
   \label{eq:ls}
 \end{align}

Similarly,  for the myopic-rationalists we have,

  \begin{align}
    \overbrace{M_s(t)}^{\substack{\text{number of }\\ \text{vaccine-seeker myopic} \\ \text{rationalists}}} = \overbrace{(\alpha_1N_n-M_v(t))}^{\substack{\text{number of}\\\text{ unvaccinated}\\ \text{myopic rationalists}}}\overbrace{H(\Delta \pi(t))}^{\substack{\text{sign indicator of} \\\text{excess payoff}}}.
     \label{eq:ms}
 \end{align}

Providing there are sufficient vaccine doses, each vaccine seeker can get vaccinated. In the presence of vaccine limitation, however, not all vaccine seekers can be inoculated at once. The available doses are then assigned randomly to the vaccine seekers of each group of decision-making types. The per capita available vaccine doses for vaccine-seekers can be formulated as  $v(t)/(L_s(t)+M_s(t))$  where $v(t)$ denotes the number of available doses at time $t$ for the first shot and $L_s(t)+M_s(t)$ represents the total demand for vaccination at time $t$. Having $v(t)/(L_s(t)+M_s(t)) > 1$ simply means that given the available doses, each vaccine seeker has the opportunity to receive the vaccine.
the rate of change of vaccinated success-based learners can be written as

\begin{align}
    \overbrace{\Dot{L}_{v}(t)}^{\substack{\text{rate of change of}\\ \text{vaccinated success} \\ \text{based learners}}} &= \overbrace{\kappa}^{\substack{\text{rate of}\\{\text{vaccination}}}}  \overbrace{L_s(t)\min \{1,\frac{v(t)}{L_s(t)+M_s(t)} \}}^{\substack{\text{number of vaccine-seeker }\\\text{success-based learners}\\\text{who can get a vaccine}}},
     \label{eq:lvdot}
   \end{align}

   where $\kappa$ represents the vaccination rate. Similarly, for the rate of change of vaccinated myopic rationalists, we have 

   \begin{align}
    \overbrace{\Dot{M}_{v}(t)}^{\substack{\text{rate of change of}\\ \text{vaccinated myopic} \\ \text{rationalists}}} &= \overbrace{\kappa}^{\substack{\text{rate of}\\{\text{vaccination}}}}  \overbrace{M_s(t)\min \{1,\frac{v(t)}{L_s(t)+M_s(t)} \}}^{\substack{\text{number of vaccine-seeker }\\\text{myopic  rationalists}\\\text{who can get a vaccine}}}.
    \label{eq:mvdot}
   \end{align}
Both types of decision-makers share the same excess payoff function \eqref{eq:excesspayoff}. The excess payoff function, however, impacts each group differently. Myopic rationalists verify whether the payoff of vaccination is higher than the that of remaining unvaccinated and if so proceed to vaccination immediately, although their vaccination rate will be affected by $\kappa$ capturing, e.g., the limitation in facilities. Contrary to myopic rationalists, success-based learners are influenced by the value of the excess payoff. 
\subsection*{Parameter Estimation}
In Proposition 1 in the SI, we proved the identifiability of all parameters 
of the equations \eqref{eq:ls}, \eqref{eq:ms}, \eqref{eq:lvdot}, and \eqref{eq:mvdot} that govern the evolution of the number of vaccinated individuals.
The parameters 
$\{\alpha_1,  c_{{v}_0}, c_{{\bar{v}}}, {C}_i, \sigma, \kappa\}$ were estimated  by fitting the derived equations to time series data. The actual effectiveness of the first dose in death prevention was estimated to be $85\%$ \cite{bernal2021effectiveness}. Its perceived effectiveness  was set equal to $100\%$ and the cost of dying from COVID-19 was chosen to be $1$ (the highest possible value), which resulted in $C_d=1$. 
For each jurisdiction, the valid interval for $C_i$ was capped at $\max_t D(t)/I(t)$ to limit the impact of morbidity on the excess payoff by that of mortality.
The valid intervals for $c_{{v}_0}$ and $ c_{{\bar{v}}}$  were then capped at $1$. 
The valid interval for $\kappa$ was set to $[0,10]$.
 The constant of proportionality, $\sigma$, was bounded at $1$.
A time-varying $c_{\bar{v}}$ was  also considered to capture possible impacts of differentiating policies based on vaccination on the perceived excess payoff of vaccination.
    In this regard, $c_{\bar{v}}$ is allowed to be varying over the time following a piece-wise-linear function--Fig S1.
    A time-varying $c_{\bar{v}}$ was captured by three additional parameters, i.e., $s_f, \eta$, and  $s_u $. 
More specifically, as of the announcement date of the differentiating policies, 
a linear increment in the perceived cost of remaining unvaccinated is introduced with a start value, $C_{\bar{v_0}}$
, inclination, $s_u$, and the peak value $\eta C_{\bar{v_0}}$,
as free parameters. 
The increment was then followed by a linear reduction 
whose inclination was a free parameter denoted by $s_f$.

The initial conditions, $L_v(0)$, $M_v(0)$ were set to zero. 
The power law exponent, $\lambda$, was determined using the available data on Americans' concern about vaccine side-effect (see SI).
We fit the model to the reported number of weekly newly  vaccinated individuals, i.e.,  ${n}_v[k]=N_v[k]-N_v[k-1]$ with ${n}_v[0] = N_v[0]$.
The error function was chosen to be the residual sum of squares, i.e., $\Sigma _{k}||{n}_{v} [k] -\hat{n}_{v}[k] ||^2$, where $\hat{n}_{v}[k]$ denotes the estimated number of those who received their first dose of COVID-19 vaccine at time $k$. 
The error was minimized using the dual annealing optimization algorithm \cite{xiang2013generalized} and its Python implementation \cite{2020SciPy-NMeth}. 
We obtained the $95\%$ confidence intervals of the estimated parameters using both parametric and non-parametric bootstrapping approaches. 
Following \cite{farcomeni2021ensemble} $500$ bootstrapped datasets were synthesized. 
For implementing the parametric bootstrap, for each time $k$, we assumed a Poisson error structure whose mean was the estimated number of newly vaccinated individuals at $k$. 
For each bootstrapped dataset, the number of newly vaccinated individuals at time $k$ was drawn from the constructed Poisson distribution for week $k$ \cite{chowell2017fitting}. To synthesize the datasets based on the non-parametric approach, due to serial correlation between residual data points, an auto-regressive model was constructed and out of the resultant uncorrelated residuals, $500$ time-series were drawn with replacements \cite{bhattacharjee2022inference}; see SI for more details. 
The $95\%$ confidence interval then was calculated using the percentile approach \cite{davison1997bootstrap}.
After inferring the parameters of the excess payoff function with a time-varying $C_{\bar{v}}$, it turned out that this imposes  high variability in the estimated parameters (Tables S17-S24), hence, we modeled $c_{\bar{v}}$ as a constant free parameter.
 In addition, it turned out that although $\sigma$ is identifiable, its point estimate is not reliable (Table S16). Hence, it is set to one.
 Fixing the constant of proportionality to one, did not, however, impact the estimated value of our parameter of interest, i.e., $\alpha_1$. More specifically, the percentage of variations in the estimated $\alpha_1$ was less than $10\%$ for all jurisdictions.
\subsection*{Correlation with Explanatory Variables}
The linear correlations between possible explanatory variables and the estimated proportion of people who behaved like myopic rationalists in taking the first dose of COVID-19 vaccine, $\alpha$ were investigated. 
There was insufficient data available for the D.C. across most potential explanatory variables. Additionally, the distribution of the proportion of myopic rationalists did not follow a normal distribution, as indicated by the Shapiro test. The distribution, however, became normal by excluding the D.C. and the state of Vermont. Consequently, these two jurisdictions were excluded from the linear regression analysis. 
For linear correlation, we used {Pearson-r coefficient} and the simple linear regression was performed using the Python implementation of Ordinary Least Squares \cite{seabold2010statsmodels}.

}

\subsection*{Acknowledgments}

\noindent \textbf{Funding:}
The project was funded by Emerging Infectious Disease Modelling Program (NSERC, CANMOD) and Brock University.

\noindent \textbf{Competing interests:}
Authors declare that they have no competing interests.

\bibliography{ref}
\bibliographystyle{ieee}
\end{document}